\documentclass[aps,prl,twocolumn,showpacs,amsmath,amssymb,superscriptaddress,letterpaper]{revtex4-1}
\usepackage[pdftex]{graphicx} 
\usepackage{dcolumn} 
\usepackage{bm} 
\usepackage{relsize}
\usepackage{amsmath}
\usepackage{color}
\usepackage[
pdfstartview=XYZ,
bookmarks=false,
colorlinks=true,
linkcolor=blue,
urlcolor=blue,
citecolor=blue,
pdftex,
bookmarks=false,
linktocpage=true, 
hyperindex=false
]{hyperref}
\usepackage{psfrag}
\usepackage{epstopdf}
\usepackage{tocloft}
\definecolor{olive}{rgb}{0,0.6,0.4}

\begin{document}

\title{Anisotropy-driven spin relaxation in germanium}
\author{Pengke Li}
\altaffiliation{These authors contributed equally to this work} 
\affiliation{Dept. of Physics and CNAM, U. of Maryland, College Park, Maryland, 20742}
\affiliation{Dept. of Electrical and Computer Engineering, U. of Rochester, Rochester, New York, 14627}
\author{Jing Li}
\altaffiliation{These authors contributed equally to this work} 
\affiliation{Dept. of Physics and CNAM, U. of Maryland, College Park, Maryland, 20742}
\author{Lan Qing}
\affiliation{Dept. of Physics and Astronomy, U. of Rochester, Rochester, New York, 14627}
\author{Hanan Dery}
\affiliation{Dept. of Electrical and Computer Engineering, U. of Rochester, Rochester, New York, 14627}
\affiliation{Dept. of Physics and Astronomy, U. of Rochester, Rochester, New York, 14627}
\author{Ian Appelbaum}
\email{appelbaum@physics.umd.edu}
\affiliation{Dept. of Physics and CNAM, U. of Maryland, College Park, Maryland, 20742}


\begin{abstract}
A unique spin depolarization mechanism, induced by the presence of g-factor anisotropy and intervalley scattering, is revealed by spin transport measurements on long-distance germanium devices in a magnetic field longitudinal to the initial spin orientation. The confluence of electron-phonon scattering (leading to Elliott-Yafet spin flips) and this previously unobserved physics enables the extraction of spin lifetime solely from spin-valve measurements, without spin precession, and in a regime of substantial electric-field-generated carrier heating. We find spin lifetimes in Ge up to several hundreds of ns at low temperature, far beyond any other available experimental results.
\end{abstract}
\maketitle

Manipulation of electron spin in semiconductors, by utilizing its coupling to a real or effective magnetic field, is fundamental to the implementation of spintronics devices \cite{Zutic_RMP2004,Fabian_APS2007,Dery_Nature07}. The primary concept is that a transverse magnetic field (perpendicular to the initial spin orientation) will rotate the spin by inducing coherent Larmor precession. Although longitudinal magnetic fields (along the initial spin orientation) are widely used to set the magnetizations of the ferromagnetic spin injectors and initialize the $\uparrow/\downarrow$ spin state, they generally do not affect the spin orientation during transport in the paramagnetic transport channel \cite{Johnson_PRL85,Crooker_Science05,Appelbaum_Nature07,Lou_NatPhys07, Fukuma_NatMater11}. 

This simple expected behavior of conduction electron spin in a magnetic field, however, is not universal in all semiconductor materials. As studied by Chazalviel \cite{Chaza_JPCS75}, in systems with conduction band valley degeneracy and spin-orbit-induced anisotropic Land\'{e} $g$-factor [such as in germanium (Ge)], an unusual effect can occur: For an electron in a valley whose axis is oriented along $\hat{z}$ at an angle $\theta$ with an external magnetic field $\vec{B}$, we can choose $\hat{x}$ such that the Zeeman Hamiltonian governing state evolution is

\begin{equation}
\mathcal{H}=\mu_B\left(  g_\parallel B\cos\theta  \sigma_z + g_\perp B \sin\theta \sigma_x  \right),
\end{equation}

\noindent where $g_\parallel$ is the g-factor for fields along the valley axis and $g_\perp$ is for those perpendicular to it, $\mu_B$ is the Bohr magneton, and $\sigma_{x,z}$ are the $2\times 2$ Pauli spin-1/2 matrices.  This seemingly trivial Hamiltonian can be algebraically transformed into an equivalent picture for a free electron with g-factor $g_0\approx 2$:

\begin{align}
\mathcal{H}&=\mu_Bg_0\frac{B}{g_0}\left( (g_\parallel-g_\perp)\cos\theta\sigma_z  +g_\perp \sin\theta \sigma_x +  g_\perp \cos\theta \sigma_z  \right) \nonumber\\
&=g_0\mu_B\left( B \frac{g_\parallel-g_\perp}{g_0} \cos\theta \sigma_z   +\frac{g_\perp}{g_0}\vec{B}\cdot\vec{\sigma}\right).\label{EQ:hamiltonian}
\end{align}

\begin{figure}[t!]
\begin{center}
\includegraphics[width=8.5cm]{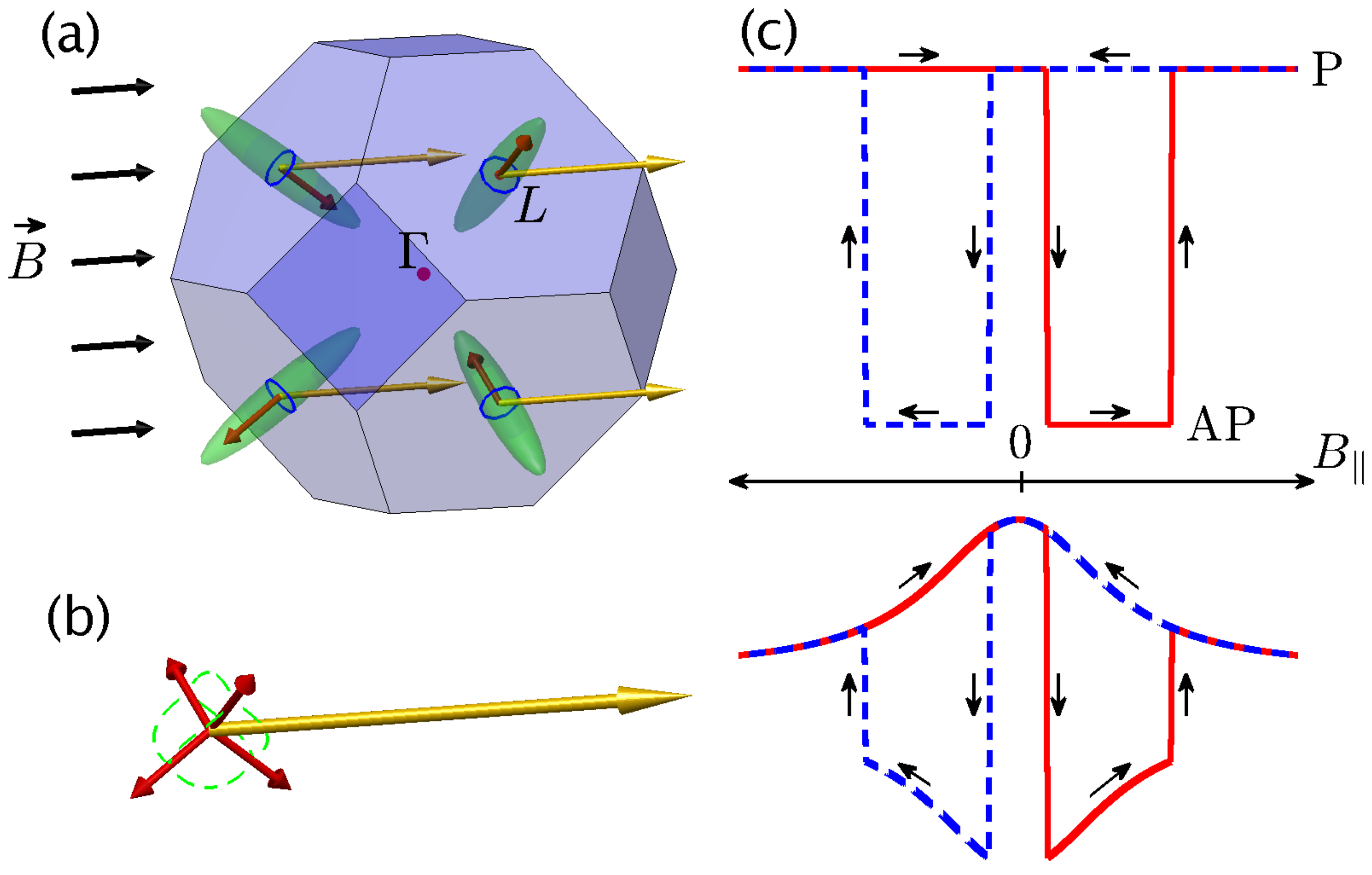}
\caption{(Color online) (a) The first Brillouin zone of germanium, showing isoenergetic surfaces of the four equivalent $L$-point conduction band valleys (green). In an external magnetic field $\vec{B}$, g-factor anisotropy causes electrons to experience an additional field oriented along the valley axis (red). (b) Equivalent magnetic field experienced by conduction electrons during intervalley scattering, where the four shorter (red) arrows represent a randomly fluctuating component. (c) Ordinary ``spin-valve'' effect in a spin transport device is shown above the $B$ axis. The red solid and blue dashed curves are, respectively, the spin signals when the magnetic field is swept from negative to positive and vice versa to orient the injector/detector magnetization configuration parallel (P, higher signal level) or anti-parallel (AP, lower signal level). Below it is the expected spin-valve effect in the presence of g-factor anisotropy and intervalley scattering, where signal decay with increasing field intensity is evident.
} \label{fig:BZ}
\end{center}
\end{figure}

Our transformed Hamiltonian shows that the electron spin acts as if it were a free electron in a renormalized magnetic field $\frac{g_\perp}{g_0}\vec{B}$, \emph{plus another magnetic field, oriented along the valley axis,} with magnitude $|B| \frac{g_\parallel-g_\perp}{g_0} \cos\theta$ [see Fig.~\ref{fig:BZ}(a)]. This additional field is randomized during the fast intervalley scattering process and thus opens a new channel of spin relaxation, even if the external magnetic field is aligned with the initial spin orientation [see Fig.~\ref{fig:BZ}(b)]. 

This extraordinary mechanism is reminiscent of the D'yakonov-Perel' spin relaxation process which dominates in non-centrosymmetric semiconductor crystal lattices \cite{Dyakonov_SPJETP1971, *Dyakonov_SPSS1972}. In that case, broken \emph{spatial inversion} symmetry allows spin-orbit interaction to cause a momentum-dependent spin splitting; \emph{intra}valley scattering during spin precession about this random effective magnetic field leads to depolarization. In the anisotropic g-factor mechanism described above, the origin of the additional random field is rooted instead in the broken \emph{time reversal} symmetry induced by the real external magnetic field, and \emph{inter}valley scattering allows g-factor anisotropy to drive its fluctuation between four different orientations. 

In the present Letter, we demonstrate this subtle phenomenon by straightforwardly showing the suppression of spin polarization with longitudinal magnetic field in the spin-valve effect [see the lower part of Fig.~\ref{fig:BZ}(c), compared with ordinary spin-valve effect above it]. To accomplish this task, we have adapted device fabrication methods to employ spin-polarized ballistic hot electron injection and detection \cite{Appelbaum_Nature07,Huang_PRL2007} in intrinsic Ge. Our control over electron transport and intervalley scattering with an independently tunable electric field reveals physics obscured due to large magnetic fields in electron resonance \cite{Roth_PR1959, Gershenzon_JETPL1970, Melnikov_SPJETP1972} and absent in optical techniques \cite{Guite_PRL11,Guite_APL12, OO_84, Hautmann_PRB12,Hautmann_PRB11, Pezzoli_PRL12} and electrical methods \cite{Liu_Nanolett10, Zhou_PRB11} due to degenerate doping conditions. Remarkably, we find spin lifetimes up to several hundreds of ns at low temperature in bulk, intrinsic Ge. These values are far beyond any other available experimental results and approach the upper limit of the intrinsic spin lifetime given by recent theory \cite{Pengke_PRB12}.

The all-electrical vertical spin-transport devices used in our experiment are nominally similar in geometry and band diagram to Si-based devices discussed in previous work \cite{Huang_APL08}. Briefly, spin-polarized hot electrons are emitted through a ferromagnetic CoFe/Al$_2$O$_3$/Al tunnel junction, travel ballistically over a Cu/Ge Schottky barrier, and then thermalize into states near the bottom of the Ge conduction band. An applied voltage ($V_{C1}$) induces electron drift through the 325~$\mu$m-thick undoped Ge (001) transport channel (room-temperature resistivity $>$40~$\Omega\cdot$cm), where magnetic fields manipulate the spin state. At the Ge/CoFe/NiFe/Cu/n-Si detector on the other side of the Ge wafer \cite{Altfeder_APL2006}, spin-dependent inelastic scattering in the ferromagnetic layers results in a signal current ($I_{C2}$) sensitive to spin polarization projected along its magnetization axis. Further device details and fabrication methods are elaborated in the Supplemental Materials \cite{SUPPLMAT}.

\begin{figure}
\begin{center}
\includegraphics[width=8.5cm]{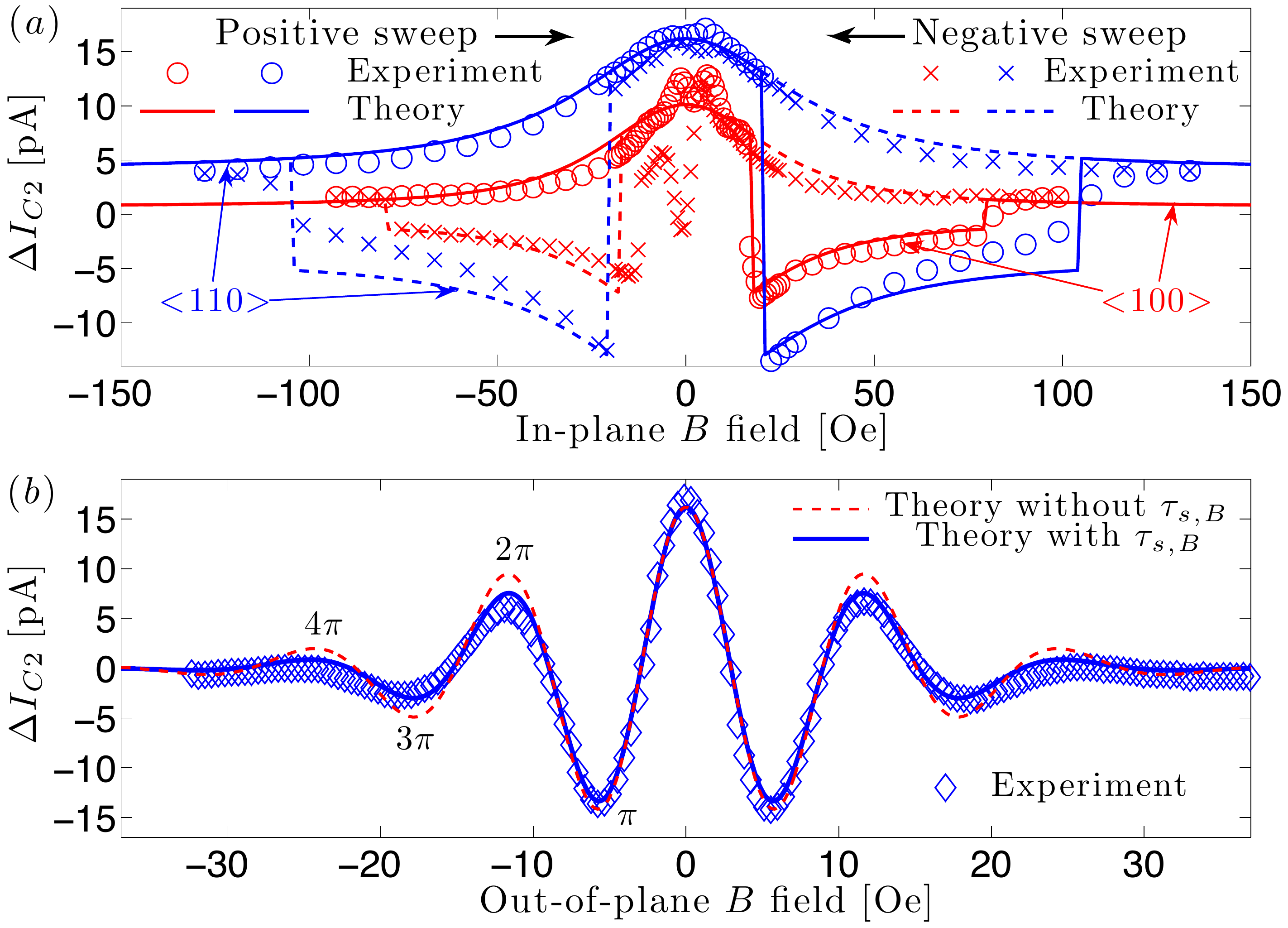}
\caption{(Color online) Experimental and simulated spin signal $\Delta I_{C2}$ vs. applied magnetic field $B$, in an accelerating electric field caused by voltage of $V_{C1}=0.6$~V over 325$\mu$m transport distance in undoped Ge at a temperature of 41~K. Panel (a) shows spin-valve effect for $\vec{B}$ field along both $<$110$>$ (in blue) and $<$100$>$ (in red) due to switching injector/detector magnetizations in an in-plane $B$ field. The round (cross) markers are experimental results when $B$ is swept in the positive (negative) direction, with the solid (dashed) curve corresponding to theoretical simulation. Panel (b) shows coherent spin precession in an out-of-plane $B$ field. The diamond markers are experiment results. The solid (dashed) curve is the theoretical simulation with (without) the g-factor anisotropy-induced depolarization [Eq.~(\ref{eq:t_total})].
} \label{fig:Bdata}
\end{center}
\end{figure}

Figure~\ref{fig:Bdata}(a) shows experiment results from these devices in an in-plane magnetic field at a temperature of 41~K. A linear background has been subtracted for clarity to show only the spin-dependent current $\Delta I_{C2}$ for magnetic field orientation along the in-plane $<$110$>$ and $<$100$>$ directions. Both data show prominent magnetic field-dependent spin depolarization with a profile very different from the ordinary spin-valve effect illustrated at the top of Fig.~\ref{fig:BZ}(c).  

This behavior cannot be accounted for only by the Elliott-Yafet (EY) spin relaxation mechanism \cite{Yafet_SSP63,Elliott_PR54} (expected to dominate in inversion-symmetric diamond-lattice materials), which involves intervalley scattering with large wavevector phonons around the Brillouin zone-edge $X$ point. In the EY process, intervalley spin-flip scattering couples the lowest and upper conduction band $L$-point components of opposite spin eigenstates \cite{Pengke_PRB12}. Such coupling is independent of the wavevector $k$, identical to intervalley momentum (spin-conserving) scattering \cite{Weinreich_PR59}. Therefore, the rates of spin relaxation ($1/\tau_{s,ph}$) and of momentum relaxation ($1/\tau_{m,ph}$) share the same temperature dependence, but with different coefficients: $\tau_{s,ph}\simeq\varrho\tau_{m,ph}$. Theoretical study \cite{Pengke_PRB12} reveals that $\varrho$ is related to the ratio of the energy gap and the strength of spin-orbit interaction between the lowest and upper conduction bands and is of the order of $10^2$. The consistent theoretical \cite{Pengke_PRB12} and experimental \cite{Weinreich_PR59} results of $\tau_{m,ph}$ in intrinsic germanium, as well as $\varrho$, give $\tau_{s,ph}\approx$ 100ns within the temperature range of 30 to 60~K. There is therefore no role for magnetic field in the bare EY mechanism in germanium.

In a magnetic field, the randomly fluctuating component aligned with the changing valley axis [identified in Eq. (\ref{EQ:hamiltonian})] provides an exponential autocorrelation function and therefore a Lorentzian power density. The spin-lattice relaxation rate for this mechanism \cite{Chaza_JPCS75, Dyakonov_SPJETP1971, *Dyakonov_SPSS1972} is  

\begin{equation}
\frac{1}{\tau_{s,B}}=\eta \,\xi\, \frac{\omega^2\, \tau_{m,ph}}{1+\omega^2\, \tau_{m,ph}^2},
\label{eq:t_B}
\end{equation}

\noindent where $\eta = (\alpha^4+\beta^4+\gamma^4)$ is between $\frac{1}{3}$$\sim$1, with $\alpha$, $\beta$ and $\gamma$ the directional cosines of $\vec{B}$ with respect to the lattice coordinates. $\xi=2\left(\frac{g_\parallel-g_\perp}{g_\parallel+2g_\perp}\right)^2\approx0.11$ relates to the $g$-factor anisotropy and $\omega=\bar{g}\mu_B B/\hbar$ is the Larmor frequency. Here, $\bar{g}=(g_\parallel+2g_\perp)/3\approx1.54$ is an averaged value due to intervalley scattering, rather than the effective $g$-factor of each valley. In the analysis of $\tau_{s,B}$, we neglect changes to momentum relaxation from cyclotron motion \cite{Wilamowski_PRB2004} due to suppression of spin relaxation up to third order in $q=k-k^\prime$ which results from the symmetry properties of the $L$-point \cite{Pengke_PRB12}. The continuous (and energetically elastic) nature of the cyclotron path in $k$-space yields an infinitesimal $q$ and therefore vanishing contribution to the overall spin-flip rate.

Including both the g-factor anisotropy contribution from external $B$ field [Eq. (\ref{eq:t_B})], and EY intervalley scattering with phonons using Matthiessen's rule, we find the total spin relaxation 

\begin{equation}
\tau_{s}\simeq \frac{1+\omega^2\, \tau_{m,ph}^2}{1+\varrho\,\eta\,\xi\, \omega^2\, \tau_{m,ph}^2}\, \tau_{s,ph}.
\label{eq:t_total}
\end{equation}

\noindent This expression remarkably shows that $\tau_{s}$ is shortened by over an order of magnitude (1/$\varrho \eta \xi$) in the pathological limit $B\rightarrow\infty$, unlike any other semiconductor system studied for spintronics.

With this $B$-dependent total spin relaxation rate, we are able to simulate the spin-valve experiment data, using a drift-diffusion model \cite{Crooker_Science05,Huang_PRB07} that takes into account the transit time uncertainty, and hence spin orientation distribution at the detector.

The EY spin lifetime $\tau_{s,ph}$ is the \emph{only} free fitting parameter in this theory, allowing us to determine relaxation rates from a single spin-valve measurement –- not possible with ordinary spin-valve measurements in materials where this mechanism is absent. The intervalley momentum lifetime $\tau_{m,ph}$, as well as the total spin lifetime $\tau_s$, are thus obtained from knowledge of $\varrho$ and Eq. (4), respectively. As can be seen by the direct comparison in Fig.~\ref{fig:Bdata}(a), the theory matches the experimental result for both $<$110$>$ and $<$100$>$ in-plane magnetic field orientations very well when $\tau_{s,ph}=$258~ns \footnote{We use a temperature-independent $\varrho=170\pm10$, obtained by the ratio of intervalley momentum scattering rate from Ref. \onlinecite{Weinreich_PR59} to spin relaxation rate from Ref. \onlinecite{Pengke_PRB12} in subsequent simulation results.}. Note that the value $\eta$ equals 0.5 and 1, respectively, for these two orientations; with a smaller $\eta$ in the denominator of Eq.~(\ref{eq:t_total}), the spin lifetime is longer and the depolarization seen in spin-valve measurements is suppressed. This long spin lifetime at 41~K is the consequence of vanishing intravalley spin flips due to time reversal and spatial inversion symmetries at the $L$-point up to cubic order in phonon wavevector \cite{Pengke_PRB12}. 

Figure~\ref{fig:Bdata}(b) shows data and the corresponding drift-diffusion simulation results for a measurement in out-of-plane magnetic field, perpendicular to the injected spin orientation, causing coherent precession and an oscillating spin detector signal. In this geometry, we must include both spin-lattice (depolarization) and spin-spin (dephasing) relaxation. The latter has the same form given by Eq.~(\ref{eq:t_B}) except that the orientation-dependent parameter is $\eta = 1-\frac{1}{2}(\alpha^4+\beta^4+\gamma^4)$ \cite{Chaza_JPCS75}. Clearly, the theoretical simulation matches experimental data very well using the same $\tau_{s,ph}=$258~ns. For comparison, we also show the simulated result excluding the g-factor anisotropy-induced contribution to the spin relaxation. Its discrepancy with experimental data is not apparent at low precession angles in small fields, but becomes prominent at subsequent extrema corresponding to 2$\pi$, 3$\pi$ and 4$\pi$ rotations when $B$ increases.

\begin{figure}
\begin{center}
\includegraphics[width=8.5cm]{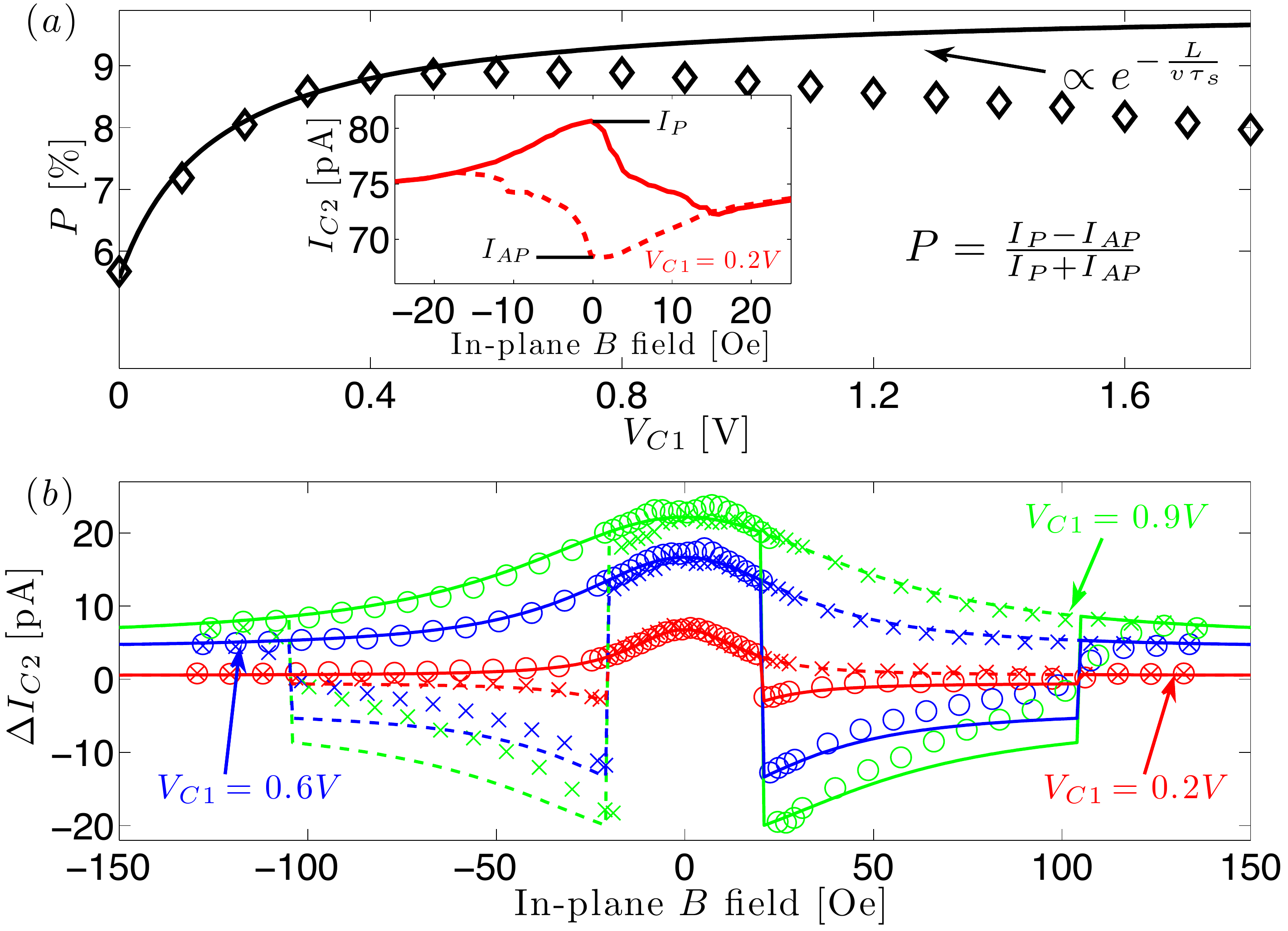}
\caption{(Color online) (a) Minor-loop-derived spin polarization at zero magnetic field as a function of accelerating voltage ($V_{C1}$) across 325$\mu$m undoped Ge spin transport channel. A typical minor loop measurement is shown in the inset. Comparison to the expected behavior from transit time depolarization with a static spin lifetime indicates electron heating and enhancement of intervalley scattering. (b) Spin valves with a $<$110$>$-oriented in-plane $B$ field for various electric fields showing evolution of the anisotropy-driven depolarization due to shorter transit times and reduction in g-factor anisotropy-induced spin lifetime.}
\label{fig:F3}
\end{center}
\end{figure}

Our independent control over electric field in the Ge transport channel allows us to change not only the transit time from injector to detector, but also the intervalley scattering rate $\tau_{m,ph}$ through Coulomb acceleration and electron heating. As shown in Fig.~\ref{fig:F3}(a), increasing the accelerating voltage $V_{C1}$ from zero at first causes an increase in spin polarization at zero magnetic field (as measured with parallel/antiparallel ($I_P$/$I_{AP}$) minor-loop in-plane measurements via $P=\frac{I_P-I_{AP}}{I_P+I_{AP}}$, shown in the inset). This low electric-field regime can be used in conjunction with transit time obtained from Fourier transform of the spin precession data (a ``Larmor clock'' method \cite{SUPPLMAT, Jang_PRL2009, Huang_PRB2010}) to extract the spin lifetime \cite{Huang_PRL2007}. However, at higher electric fields (above $V_{C1}/L\approx$0.6~V/325~$\mu$m$\approx$20~V/cm), the measured spin polarization begins to \emph{decrease}, which is inconsistent with a static spin lifetime and positive differential mobility that would yield a polarization $\propto e^{-\frac{L}{v\tau_s}}$, valid here in drift-dominated transport when $\sqrt{\frac{D}{Lv}}\ll 1$, where $D$ is diffusion coefficient and the drift velocity $v$ is a monotonic function of electric field. This ``negative differential spin lifetime'' behavior was first seen in spin transport experiments with intrinsic Si, but at much higher electric fields $\approx$3~kV/cm \cite{Li_PRL12}. It similarly results from an increase in electron kinetic energy and consequently an enhanced intervalley scattering rate; however, the intervalley spin-flip probability in Ge is an order of magnitude larger than that of the $f$-process in Si \cite{Song_PRB2012, Pengke_PRB12}, and much greater mobility in Ge causes a higher electron temperature at the same electric field \cite{Jacoboni_PRB1981}.

Here with Ge, we can use the spin-valve features to determine the spin lifetime in this ``warm electron'' regime, a task that is not straightforward in Si without sophisticated Monte-Carlo calculations \cite{Li_PRL12} because (although it has strong intervalley spin depolarization) its $\Delta$-axis conduction valleys have negligible g-factor anisotropy \cite{Roth_PR1960}. Figure~\ref{fig:F3}(b) shows how this signal in Ge evolves with electric field: At low voltage ($V_{C1}=$0.2~V), the transit time is relatively long and intervalley scattering is determined by the density of phonons at equilibrium with the lattice temperature. At higher voltage ($V_{C1}=$0.9~V), the transit time drops but the magnetic field fluctuation frequency increases due to increase in intervalley scattering rate. Simulations using Eq.~(\ref{eq:t_total}) fit our empirical data with $\tau_s=$  393$\pm$33~ns for $V_{C1}=$0.2~V, 258$\pm$29~ns for 0.6~V and 167$\pm28$~ns for 0.9~V, showing a strong electron heating-induced spin depolarization.  

\begin{figure}
\begin{center}
\includegraphics[width=8.5cm]{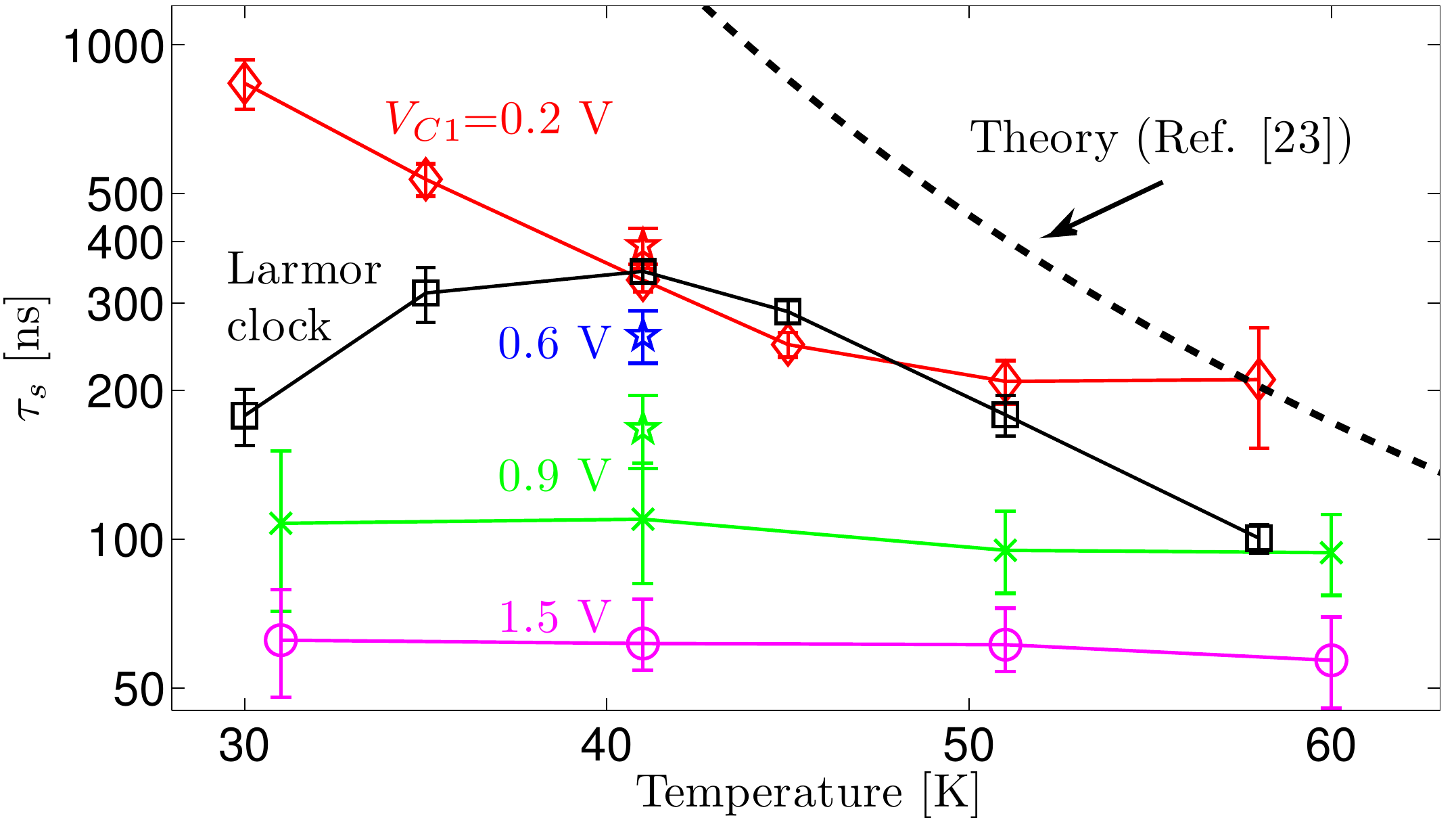}
\caption{(Color online) Temperature dependence of spin lifetime in undoped Ge. Low-field measurements at $V_{C1}=$0.2~V yield a monotonically increasing spin lifetime with decreasing temperature, similar to the Elliott-Yafet theoretical prediction from Ref. \onlinecite{Pengke_PRB12}.  Electric-field-induced intervalley scattering causes enhanced suppression of lifetimes extracted from fitting the effects of Eq.~(\ref{eq:t_total}) to spin-valve measurement in a $<$110$>$-oriented longitudinal magnetic field at accelerating voltages $V_{C1}$=0.9 and 1.5V. For comparison, lifetime obtained by correlating transit time (``Larmor clock'' from spin precession measurements) and zero-magnetic-field polarization (from minor loop spin-valve measurements) show low-temperature depolarization inconsistent with Ref. \onlinecite{Pengke_PRB12}.}
\label{fig:F4}
\end{center}
\end{figure}

Because the electron temperature is easily decoupled from the lattice temperature in transport conditions at finite electric field in this material, the spin lifetimes extracted from fitting the spin-valve depolarization features are typically lower than those obtained by correlating zero-magnetic-field polarization with mean transit time from spin precession data \cite{Huang_PRL2007, Li_PRL12} except at the lowest accelerating voltages and temperatures. Figure~\ref{fig:F4} compares the temperature dependence of spin lifetimes at several internal electric fields to the theoretical EY prediction from Ref. \onlinecite{Pengke_PRB12}, which applies to the germanium electron-phonon system at thermal equilibrium. The spin-valve-obtained lifetimes systematically drop with increasing electric field, and are noticably temperature independent for high $V_{C1}$, unlike the Larmor-clock-derived values from fitting precession and minor-loop spin polarization data at $V_{C1}<$0.6~V. The origin of low-temperature spin lifetime suppression seen in these data is likely due to extrinsic effects, as has been observed in electron spin resonance studies of Si \cite{Lepine_PRB1970}. 

In closing, we remark that although our observation of anisotropy-driven spin relaxation is confined to the present work on elemental Ge, we do not expect it to be unique in this regard. Many compound semiconductors with the inversion-symmetric rocksalt crystal lattice (heretofore overlooked as spin transport materials) have both strong and anisotropic spin-orbit interaction, and equivalent conduction states at the $L$-point. If the quantity $\xi=2\left(\frac{g_\parallel-g_\perp}{g_\parallel+2g_\perp}\right)^2$ in these materials is comparable to its value in Ge ($\approx$0.1), then we expect to see similar features in the spin-valve behavior when intervalley scattering is induced by finite carrier temperature. Regardless of the material, this phenomena can be suppressed by quenching intervalley scattering with strain oriented to isolate one conduction band valley \cite{Gershenzon_SPS1976}.

\begin{acknowledgments}
We acknowledge the support of the Maryland NanoCenter and its FabLab. This work was supported at U. of Maryland by the Office of Naval Research under contract N000141110637, the National Science Foundation under contracts ECCS-0901941 and ECCS-1231855, and the Defense Threat Reduction Agency under contract HDTRA1-13-1-0013. Work at U. of Rochester was supported by NSF contracts DMR-1124601 and ECCS-1231570, and by DTRA contract HDTRA1-13-1-0013.  
\end{acknowledgments}

%

\end{document}